\documentclass{ar-1col}
\usepackage{url}
\usepackage{natbib}
\usepackage{ulem}
\usepackage{natbib}

\jname{Annu. Rev. Fluid Mech.}
\jvol{50}
\jyear{2018}
\doi{10.1146/annurev-fluid-122316-044539}

\begin{document}

\markboth{Dauxois et al.}{Instabilities of internal wave beams}

\title{Instabilities of Internal Gravity Wave Beams}

\author{Thierry Dauxois, Sylvain Joubaud, Philippe Odier and Antoine Venaille
\affil{Univ Lyon, ENS de Lyon, Univ Claude Bernard, CNRS, Laboratoire de Physique, F-69342 Lyon, France; email: Thierry.Dauxois@ens-lyon.fr}}

\begin{abstract}

Internal gravity waves play a primary role in geophysical fluids:  they contribute significantly to mixing in the ocean and they redistribute energy and momentum in the middle atmosphere. Until recently, most studies were focused on plane wave solutions. However, these solutions are not a satisfactory description of most geophysical manifestations of internal gravity waves, and it is now recognized that internal wave beams with a confined profile are ubiquitous in the geophysical context.

We will discuss the reason for the ubiquity of wave beams in stratified fluids, related to the fact that  they are solutions of the nonlinear governing equations.
We will focus more specifically on situations with a constant buoyancy frequency.
Moreover, in light of recent experimental and analytical studies of internal gravity beams, it is timely to discuss the two main mechanisms of instability for those beams. i) The Triadic Resonant Instability  generating two secondary wave beams. ii) The streaming instability corresponding to the spontaneous generation of a mean flow.


\today

\end{abstract}

\maketitle

\tableofcontents

\section{INTRODUCTION}

Internal gravity waves play a primary role in geophysical fluids~\citep{SutherlandBook}:  they contribute significantly to mixing in the ocean \citep{wunsch2004} and they redistribute energy and momentum in the middle atmosphere~\citep{fritts2003}.  The generation and propagation mechanisms are fairly well understood, as for instance in the case of oceanic tidal flows \citep{GarrettKunze2007}.  By contrast, the  dissipation mechanisms, together with the understanding of observed energy spectra resulting from nonlinear interactions between those waves, are still debated \citep{Johnstonetal2003,MacKinnonWinters2005,RainvillePinkel2006,CalliesFerrariBuhler,Alford2015,SarkarScotti2016}. Several routes towards dissipation have been identified, from wave-mean flow interactions to cascade processes, but this remains a fairly open subject from both theoretical~\citep{Craik,NazarenkoBook} and experimental points of view~\citep{StaquetSommeria2002}. The objective of this review is to present important recent progress that sheds new light on the nonlinear destabilization of internal wave beams, bridging part of the gap between our understanding of their generation mechanisms based mostly on linear analysis, and their subsequent evolution through nonlinear effects.

Until recently, most studies were focused on plane wave solutions, which are introduced in classical textbooks~\citep{GillBook}. Strikingly, such plane waves are not only solutions of the linearized dynamics, but also of the nonlinear equations \citep{McEwan1973,Akylas2003}. However, spatially and temporally monochromatic internal wave trains are not  a satisfactory description of most geophysical internal gravity waves~\citep{Sutherland2013}. Indeed, oceanic field observations have rather reported internal gravity beams with a confined profile \citep{LienGregg2001,Coleetal2009,Johnstonetal2011}. In the atmosphere, gravity waves due to thunderstorms also often form beam-like structures  \citep{Alexander2003}. Oceanic wave beams  arise from the interaction of the barotropic tide with sea-floor topography, as has been recently studied theoretically and numerically \citep{Khatiwala2003,Lamb2004,MaugeGerkema}, taking into account transient, finite-depth and nonlinear effects, ignored in the earlier seminal work by \cite{Bell1975}. The importance of those beams has also been emphasized recently in quantitative laboratory experiments~\citep{GostiauxDauxois2007,ZhankKingSwinney2007,PeacockEcheverriBalmforth2008}. From these different works, it is now recognized that internal wave beams are ubiquitous in the geophysical context.

The interest for internal gravity beams resonates with the usual pedagogical introduction to internal waves,
the Saint Andrew's cross, which comprises four beams generated by oscillating a cylinder in a stratified fluid~\citep{MowbrayRarity1967}. Thorough studies of internal wave beams can be found in ~\cite{voisin2003}. Moreover, \cite{Akylas2003} have realized that an inviscid uniformly stratified Boussinesq fluid 
supports time-harmonic plane waves invariant in one transverse horizontal direction, propagating along a direction determined by the frequency {(and the medium through the buoyancy frequency)}, with a general spatial profile in the cross-beam direction.  
These  wave beams are not only fundamental to the linearized dynamics but, like sinusoidal wavetrains, happen
 to be exact solutions of the nonlinear governing equations. Remarkably,  \cite{Akylas2003} showed that the steady-state similarity linear solution
for a  viscous beam  \citep{ThomasStevenson1972} is also valid in the nonlinear regime.
In light of the recent experimental and analytical studies of those internal gravity {wave} beams, 
it is thus timely to study their stability properties.

The structure of the review is the following. First, in section~\ref{sectionIntroductive}, we introduce the  subject  by presenting concepts, governing equations and approximations that lead to the description of gravity waves in stratified fluids. We dedicate a special emphasis on the peculiar role of nonlinearities to explain why internal gravity wave beams are ubiquitous solutions in oceans and middle atmospheres. Then, in section~\ref{TriadicResonanceInstability}, we discuss the classic Triadic Resonant Instability that corresponds to the destabilization  of a primary wave with the spontaneous emission of two secondary waves, of lower frequencies and different wave vectors. In addition to the simple case of plane waves, we discuss in detail  the generalization to wave beams with a finite width.  Section~\ref{StreamingInstability} is dedicated to the streaming instability, the second important mechanism for the instability of internal gravity waves beams through the generation of a mean flow.  Finally, in section~\ref{ConclusionsPerspectives}, we draw some conclusions and discuss main future issues.

\section{THE DYNAMICS OF STRATIFIED FLUIDS AND ITS SOLUTIONS}\label{sectionIntroductive}

\subsection{Basic Equations}

Let us consider an incompressible non rotating stratified Boussinesq fluid in Cartesian coordinates (\mbox{\boldmath $e$}$_x$,\mbox{\boldmath $e$}$_y$,\mbox{\boldmath $e$}$_z$) where \mbox{\boldmath $e$}$_z$ is the direction opposite to gravity. The Boussinesq approximation amounts to neglecting density variations with respect to a constant reference density $\rho_{\mathrm{ref}}$, except when those variations are associated with the gravity term $g$.  The relevant field  to describe the effect of density variations is then the buoyancy field 
$b_{\mathrm{tot}}=g\left(\rho_{\mathrm{ref}}-\rho\right)/\rho_{\mathrm{ref}}$, with $\rho(\mathbf{r},t)$ the full density field, \mbox{\boldmath $r$}=($x$,$y$,$z$) the space coordinates and $t$ the time coordinate. Let us call $\rho_0(z)$  the density of the flow at rest, with buoyancy frequency $N(z)=(-g\left(\partial_z \rho_0\right)/\rho_{\mathrm{ref}})^{1/2}$. The corresponding buoyancy profile $g\left(\rho_{\mathrm{ref}}-\rho_0\right)/\rho_{\mathrm{ref}}$ is denoted $b_0$. The buoyancy frequency $N$ varies in principle with the depth~$z$. In the ocean, $N$ is rather large in the thermocline and weaker in the abyss. For the sake of simplicity, however, $N$ will be taken constant
in the remainder of the paper. In some studies, to ease greatly the theoretical
analysis, this approximation that looks drastic  
at first sight can be relaxed
when $N$ changes smoothly by relying on the WKB approximation. 

The equations of motion can be written as a dynamical system for the perturbed buoyancy field  $b=b_{\mathrm{tot}}-b_0$ and the three components of the velocity field \mbox{\boldmath $u$} = ($u_x$,$u_y$,$u_z$):
\begin{eqnarray}
 \nabla\cdot \mbox{\boldmath $u$} &=& 0, \label{eq:div_u}\\
\partial_t \mbox{\boldmath $u$} + \mbox{\boldmath $u$}\cdot \nabla \mbox{\boldmath $u$}&=& -\frac{1}{\rho_{\mathrm{ref}}}\nabla p +b \mbox{\boldmath $e$}_z + \nu \nabla^2 \mbox{\boldmath $u$}, \label{eq:NS_strat}\\
\partial_t b + \mbox{\boldmath $u$}\cdot \nabla b +u_z N^2 &=&0  . \label{eq:cons_masse}
\end{eqnarray}
with  $p(\mbox{\boldmath $r$},t)$ the pressure variation with respect to the hydrostatic equilibrium pressure  $P_0(z)=P_{0}(0)-\int_{0}^z \rho_{0}(z') g \mathrm{d} z'$, and $\nu$ the kinematic viscosity.  We have  neglected the molecular  diffusivity, which would imply a term  $D\nabla^2 b$ in the right-hand side of Equation~(\ref{eq:cons_masse}), with $D$ the diffusion coefficient of the stratifying element (molecular diffusivity for salt, thermal diffusivity for temperature). The importance of the dissipative terms with respect to the nonlinear ones are described by the Reynolds $UL/\nu$ and the Peclet numbers $UL/D$, with $U$ and $L$ typical velocity and length scales, or equivalently by the Reynolds number and the Schmidt number $\nu/D$. In many geophysical situations, both Reynolds and Peclet numbers are large, and molecular effects can be neglected at lowest order. In such cases, the results do not depend on the Schmidt number.  In laboratory settings, the Peclet is often also very large, at least when the stratification agent is salt, in which case $D\approx10^{-9}$ m$^2\cdot$s$^{-1}$. However, the viscosity of water is  $\nu\approx 10^{-6}$ m$^2\cdot$s$^{-1}$, and the corresponding Reynolds numbers are such that viscous effects can play an important role, as we will see later.

Let us first consider the simplest case of two-dimensional flow, which is invariant in the transverse $y$-direction. The non-divergent two-dimensional velocity field is then conveniently expressed in  terms of a streamfunction $\psi(x,z)$ as  $\mbox{\boldmath $u$}=(\partial_z\psi,0,-\partial_x\psi)$. Introducing the Jacobian $J(\psi,b)=\partial_x \psi\, \partial _zb - \partial_x b\, \partial _z \psi$, the  dynamical system (\ref{eq:div_u}), (\ref{eq:NS_strat}) and (\ref{eq:cons_masse}) is expressed as 
\begin{eqnarray}
\partial_{t}\nabla^2 \psi + J(\nabla^2 \psi , \psi) &=&  -\partial_x b+\nu \nabla^4 \psi,\label{equationenpsi} \\
 \partial_t b+ J(b,\psi) - {N^2 }\partial_x \psi &=& 0. \label{equationenrho}
\end{eqnarray}
Differentiating Equation~(\ref{equationenpsi}) with respect to time and Equation~(\ref{equationenrho}) with respect to the spatial variable $x$, and subtracting the latter from the former, one gets finally
\begin{eqnarray}
\partial_{tt}\nabla^2 \psi +N^2 \partial_{xx} \psi  &=& \nu \nabla^4 \partial_t\psi + \partial_t J( \psi , \nabla^2 \psi)  + \partial_x J(b,\psi),\label{equationenpsietrho}
\end{eqnarray}
 describing the nonlinear dynamics of non-rotating non-diffusive viscous stratified fluids in two dimensions. 

\subsection{Linear Approximation}

In the linear  approximation, assuming vanishing viscosity,  the right-hand side of Equation~(\ref{equationenpsietrho})  immediately vanishes leading to the following wave equation for the streamfunction
\begin{eqnarray}
 \partial_{tt}\nabla^2 \psi + N^2 \partial_{xx}\psi &=& 0. \label{eq_disp_gravity_non_viscous}
\end{eqnarray}
This equation is striking for several reasons. First, its mathematical structure is clearly different from the traditional d'Alembert equation. Indeed, the spatial differentiation appears at second order in both terms.
Time-harmonic plane waves with frequency $\omega$, wave vector $\mbox{\boldmath $k$}=(\ell,0,m)$ and wavenumber $k=|\mbox{\boldmath $k$}|=(\ell^2+m^2)^{1/2}$ are solutions of Equation~(\ref{eq_disp_gravity_non_viscous}), if the dispersion relation for internal gravity waves 
\begin{eqnarray}
 \omega=\pm N \frac{\ell}{k} = \pm N \sin \theta, \label{eq_disp_gravity_theta}
\end{eqnarray}
is satisfied. $\theta$ is the angle between wavenumber $\mbox{\boldmath $k$}$ and the vertical.
\begin{marginnote}[]
\entry{Plane wave}{ 
$  \psi_0\, e^{i\left(\mbox{\boldmath $k$}\cdot\mbox{\boldmath $r$} -\omega t\right)}+\textrm{c.c.}$ where c.c. denotes complex conjugate}
\end{marginnote}

The second important remark is that contrary to the usual concentric waves emitted from the source of excitation when
considering the d'Alembert equation, here four different directions of propagation
are possible depending on the sign of $\ell$ and~$m$. This is 
an illustration of the anisotropic propagation due to the vertical stratification.

The third remarkable property is that the dispersion relation
features the angle of propagation rather than the wavelength, emphasizing a clear difference between internal waves
and surface waves. This is also a crucial property for this review since it will allow us to define beams with a general profile,
rather than with a single wavenumber.

\subsection{Nonlinear Terms}\label{NLterms}
\subsubsection{Plane Wave Solutions}

It is striking and pretty unusual that 
plane waves are  solutions of the {inviscid} nonlinear equation~(\ref{equationenpsietrho}) even for large amplitudes.
Indeed, the streamfunction of the plane wave solution is a Laplacian eigenmode, with $\nabla^2 \psi=-k^2\psi$. Consequently, the first Jacobian term  vanishes in  Equation~(\ref{equationenpsietrho}).
Equation~(\ref{equationenpsi})
  leads {therefore} to the 
so-called polarization relation $b=-\left( N^2\ell /\omega\right) \psi \equiv{\cal P}\psi$,  with ${\cal P}$ the polarization prefactor.
Consequently, the second Jacobian in (\ref{equationenpsietrho}) vanishes: $J(\psi,{\cal P}\psi)=0$. To conclude, 
both nonlinear terms in Equation~(\ref{equationenpsietrho}) vanish for plane wave solutions,
that are therefore solutions of the nonlinear equation, for any amplitude.

\subsubsection{Internal Wave Beams}

\begin{figure}[h]
\includegraphics[width=0.7\textwidth]{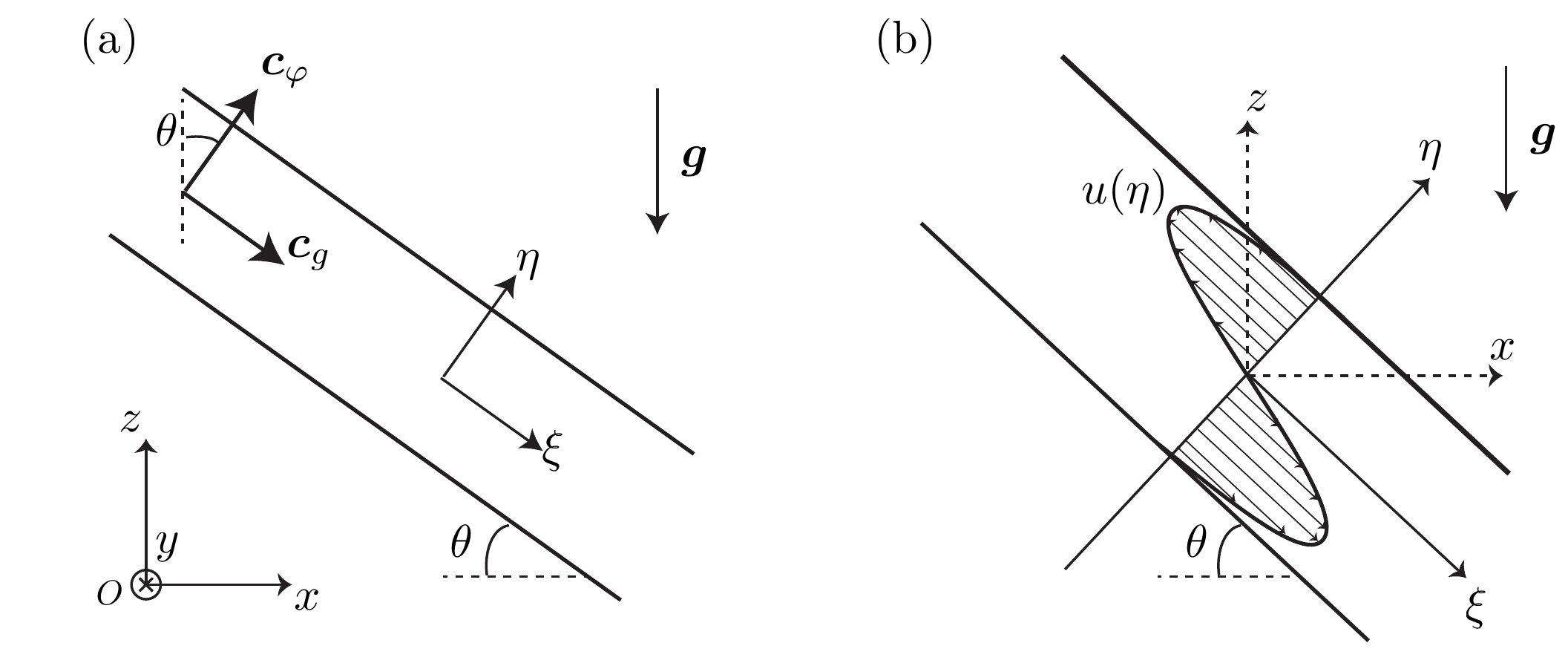} 
\vskip -.5truecm
\caption{
(a) Schematic representation of an internal wave beam and definition of the longitudinal
and cross-beam coordinates $\xi$ and $\eta$, of the angle of inclination $\theta$,
and finally of the group and phase velocities $c_g$ and~$c_\varphi$. (b) Geometry of a uniform  (along~$\xi$)  internal wave beam inclined at an angle $\theta$ with respect to the horizontal.
The beam profile varies in the cross-beam $\eta$ direction, and the associated flow velocity
 is in the along-beam direction~$\xi$. The transverse horizontal direction is denoted
by $y$. 
}
\label{profilselonetabb}
\end{figure}

{Since} the frequency $\omega$ is independent of the wavenumber, 
it is possible to devise more general solutions, time-harmonic with the same frequency~$\omega$, by superposing several linear solutions
associated to the same angle of propagation, but with different wavenumbers $k$~\citep{McEwan1973,Akylas2003}.
Introducing the along-beam coordinate $\xi = x \cos \theta - z \sin \theta$, defined along the direction of propagation,  and the cross-beam coordinate $\eta = x \sin \theta + z \cos \theta$ (see \textbf{Figure~\ref{profilselonetabb}}),
the plane wave solution  can be written as
\begin{eqnarray}
 \psi(x,y,z,t) &=& \psi_0 \, e^{i(\ell x+mz- \omega t)}+\textrm{c.c.}=
 \psi_0\, e^{ik \eta }\,e^{-i \omega t}+\textrm{c.c.}\,,\label{ondesplanesbis}
\end{eqnarray}
since $\ell=k \sin\theta$ and $m=k\cos\theta$.
If one introduces $Q(\eta)=ik \psi_0 e^{ik\eta }$, one obtains the velocity field 
$\mbox{\boldmath $u$}= Q(\eta) (\cos\theta ,{0},-\sin \theta) e^{-i\omega t } +{c.c.}$ and  the buoyancy perturbation
$b=-i({{\cal P}}{/k})Q(\eta)  e^{-i\omega t } +{c.c.}\,.
$

One can actually obtain a wider class of solutions by considering an arbitrary complex amplitude $Q(\eta)$. 
Indeed, the fields  $\mbox{\boldmath $u$}$ and $b$ do not depend on the longitudinal variable  $\xi$. 
Consequently, after the change of variables, the Jacobians, which read {$J(\psi,b)=\partial_\xi \psi\, \partial _\eta b - \partial_\xi b\, \partial_\eta \psi$}, simply vanish, 
 making the governing equations linear. 
 As discussed in~\cite{Akylas2005}, 
 note that uni-directional beams, in which energy propagates in one direction, involve plane waves with wavenumbers of the same sign only: $Q(\eta)=\int_0^{+\infty}A(k)e^{ik\eta} \mbox{d}k$ or $Q(\eta)=\int_{-\infty}^0A(k)e^{ik\eta} \mbox{d}k$.

\begin{marginnote}[]
\entry{Internal wave beam}{Superposition of  time-harmonic plane waves with an arbitrary profile in the cross-beam direction. \\ We will call uniform beam the special case of a internal plane  wave with a confined spatial profile.}
\end{marginnote}
 
We see that the class of propagating waves that are solutions of the nonlinear dynamics in a Boussinesq stratified fluid is much more general than plane wave solutions: there is a whole family of solutions corresponding to uniform plane waves in the longitudinal  direction~$\xi$, but with a general profile in the cross-beam direction~$\eta$, as represented in  \textbf{Figure~\ref{profilselonetabb}}. 

\cite{Akylas2003}  have generalized those results by computing  asymptotic solutions for a slightly viscous nonlinear
wave beam with amplitude slowly modulated along $\xi$ and in time. 
After considerable manipulation, it turns out that all leading-order nonlinear 
{advective}-acceleration terms in the governing equations
of motion vanish, and 
a uniform (along~$\xi$) beam, regardless of its profile (along~$\eta$),
represents an exact nonlinear solution in an unbounded, inviscid, uniformly stratified
fluid.  This result not only extends 
the validity of the~\cite{ThomasStevenson1972} steady-state similarity solution 
to the nonlinear regime, but emphasizes how nonlinearity has only relatively weak consequences.
This has profound and useful outcomes on the applicability of 
results obtained with linear theory, for comparisons with field observations, 
laboratory experiments or numerical simulations. 
 
The vanishing of the nonlinear contributions is really unexpected and 
results from the combination of numerous different terms. \cite{Akylas2003} noticed, however, that 
the underlying reason for the {seemingly}  miraculous cancellation of the resonant nonlinear terms 
was the very same one that had been already pointed out by \cite{DauxoisYoung1999}.
After lengthy calculations, in both cases, the reason 
is a
special case of the Jacobi identity $J\left[A,J(B,C) \right]+J\left[C,J(A,B) \right] +J\left[J(A,C),B \right]=0$.
\cite{DauxoisYoung1999} 
were studying near-critical reflection of a finite amplitude internal wave on a slope
to heal the singularity occurring in the solution of~\cite{Phillips1966}.
Using matched asymptotic, they took a
distinguished limit in which the amplitude of the incident wave, the
dissipation, and the departure from criticality are all small. 
At the end, although the reconstructed fields do contain nonlinearly driven
second harmonics,
they obtained the striking and unusual result that the final amplitude equation happens to be a linear equation. The underlying reason was already this Jacobi identity.\footnote{Studying the mechanism of superharmonic generation, \cite{Alam2016} reported recently another situation for which
the nonlinear terms vanish in the domain bulk. Interestingly, however, they play a pivotal role through the free surface boundary condition.}

\bigskip

To conclude, the effects of nonlinearities on plane waves or wavebeams exhibit very peculiar properties. 
There are two important points to keep in mind. 
First, plane waves and internal wave beams are solutions
of the full equation.
Second, identifying a solution does not mean that it is a stable one. 
 This remark is at the core of the present review: we will  focus in the following on the behavior of wave beams with respect
to the triadic resonant and the streaming instabilities.

\section{TRIADIC RESONANT INSTABILITY}\label{TriadicResonanceInstability}

\subsection{Introduction}

It was first realized fifty years ago that internal gravity plane waves are unstable to infinitesimal perturbations, 
which grow to form temporal and spatial resonant triads~\citep{
DavisAcrivos1967,
McEwan1971,
Mied1976
}.  This  nonlinear instability produces two secondary waves that extract 
energy from a primary one. Energy transfer rates due to this instability are now well established for plane waves~\citep{StaquetSommeria2002}.

The instability {was} observed in several laboratory experiments~\citep{BenielliSommeria1998,ClarkSutherland2010,Pairaudetal2010,Joubaudetal2012} and numerical experiments on propagating internal waves~\citep{Koudella2006,Wienkers2015} or reflecting internal tides
on a horizontal or sloping boundary~\citep{GerkemaStaquetBouruet-Aubertot2006,Pairaudetal2010,ZhouDiamessis2013,GayenSarkar2013}.  Oceanic field observations have also confirmed the importance of this instability, especially close to the critical latitude, where the Coriolis frequency is half of the  tidal frequency~\citep{HibiyaNagasawaNiwa2002,MacKinnonetal2013,Sun2013}.

Recent experiments by \cite{BDJO2013}, however, followed by a simple model and numerical simulations by \cite{BSDBOJ2014} as well as a theory by \cite{Karimi2014} have shown that finite-width internal gravity wave beams exhibit a much more complex behavior than expected in the case of interacting plane waves. This is what will be discussed  in this section.

\begin{textbox}[h]\section{The Triadic Resonant Instability (TRI)
 versus the  Parametric Subharmonic Instability (PSI)}
The classic Triadic Resonant Instability corresponds to the destabilization 
of a primary wave through the spontaneous emission of two secondary 
waves. The frequencies and wave vectors of these three waves 
are related by the spatial,
$\mbox{\boldmath $k$}_0 = \mbox{\boldmath $k$}_++\mbox{\boldmath $k$}_-$, 
 and the temporal, $
\omega_0 = \omega_++\omega_-$, 
resonance conditions, where the indices 0 and $\pm$ refer respectively to the primary and secondary waves.

In the inviscid case, the most unstable triad corresponds to antiparallel, infinitely long secondary wave vectors  associated with frequencies which are both half of the primary wave frequency: $ \omega_+\simeq\omega_-\simeq\omega_0/2$. Because of the direct analogy with the parametric oscillator, this particular case defines the Parametric Subharmonic Instability (PSI). This special case applies to many geophysical situations, and especially for oceanic applications.

In laboratory experiments,  viscosity plays an important role and the two secondary wave frequencies are different. 
By abuse of language, some authors have sometimes extended the use of the name PSI to cases for which secondary waves do not oscillate at half the forcing frequency.
To avoid confusion, in the general case, it is presumably more appropriate to use the acronym~TRI. 
\end{textbox}

\subsection{The simplest case of Plane Waves Solutions}

\subsubsection{Derivation of the Equations and  Plane Waves Solutions}\label{derivEquPlaneWaves}

Looking for solutions of the basic equations~(\ref{equationenpsi}) 
and~(\ref{equationenrho}) as sum of three plane waves as follows 
$b=\sum_{j}^{} R_j(t) e^{i \left(\mbox{\boldmath $k$}_j \cdot \mbox{\boldmath $r$} - \omega_j t\right)} +c.c.$
and
$\psi =\sum_{j}^{} \Psi_j(t) e^{i \left(\mbox{\boldmath $k$}_j \cdot \mbox{\boldmath $r$} - \omega_j t\right)} +c.c.$, {with $j=0$ for the primary wave}
and $j=\pm$  for the  secondary ones,
 and denoting $\dot R$  the derivative of the amplitude $R$, one gets
(see for example~\cite{Hasselman1967})
\begin{eqnarray}
\sum_{j}^{}[- k_j^2 (\dot \Psi_j - i \omega_j \Psi_j) + i \ell_j R_j - \nu k_j ^4 \Psi_j] e^{i \left(\mbox{\boldmath $k$}_j \cdot \mbox{\boldmath $r$} - \omega_j t\right)} +c.c. &=& - J(\nabla^2\psi, \psi)\,.\label{Eqpsi2}\\
\sum_{j}^{} [\dot R_j - i \omega_j R_j - i N^2 \ell_j \Psi_j] e^{i \left(\mbox{\boldmath $k$}_j \cdot \mbox{\boldmath $r$} - \omega_j t\right)} + c.c. &=& -J(b, \psi)\,,\label{Eqrho2}
\end{eqnarray}
The left-hand sides represent the linear parts of the dynamics. 
Neglecting the nonlinear terms, as well as the viscous terms and the temporal evolution of the amplitudes, one recovers the polarization expression $R_j = -{(N^2 \ell_j}/{\omega_j}) \Psi_j$  
and the dispersion relation  $\omega_j=N |\ell_j|/\sqrt{\ell_j^2+m_j^2} $.
This linear system is {resonantly} forced  by the Jacobian nonlinear terms on the right-hand side  {when} the waves fulfill\ a spatial resonance condition
\begin{equation}
\mbox{\boldmath $k$}_0 = \mbox{\boldmath $k$}_++\mbox{\boldmath $k$}_- \label{spatialcondition}
\end{equation}
and a temporal resonance condition
\begin{equation}
\omega_0 = \omega_++\omega_-\,. \label{temporalcondition}
\end{equation}
 The Jacobian terms in Equations (\ref{Eqpsi2}) and (\ref{Eqrho2}) can then be written as the sum of a resonant term that will drive the instability,  plus {some} unimportant non resonant terms.  Introducing this result into Equation~(\ref{Eqpsi2}), one obtains  three  relations between $\Psi_{{j}}$ and $R_{{j}}$ for each mode $\exp[{i(\mbox{\boldmath $k$}_{{j}} \cdot \mbox{\boldmath $r$} - \omega_{{j}} t)}]$   with  ${j} = 0, +$~or~$-$. 
 One gets 
 \begin{eqnarray}
R_\pm &=&\frac{1}{i\ell_\pm} \left[ k_\pm^2(\dot \Psi_\pm - i\omega_\pm\Psi_\pm) + \nu k_\pm^4\Psi_\pm +\alpha_\pm \Psi_0 \Psi^*_\mp\right]\,, \label{eq:Rpm}\end{eqnarray}
where 
$\alpha_\pm = (\ell_0 m_\mp - m_0 \ell_\mp) (k_0^2 - k_\mp^2)$. 
Here, one traditionally uses the ``pump-wave'' approximation, which assumes that over the initial critical growth period of the secondary waves, the primary wave amplitude, $\Psi_0$, remains constant {and that  the amplitude varies slowly with respect to the period of the wave ($\dot \Psi_j\ll\omega_j\Psi_j$).}
Differentiating the polarization expression, cumbersome but straightforward calculations~\citep{BDJO2013}
lead to first order  to  \begin{eqnarray}
\frac{{\rm d}\Psi_\pm}{{\rm d}t} & =&
|I_\pm|\Psi_0\Psi_\mp^*-\frac{\nu}{2} k_\pm^2\Psi_\pm ,\label{equation1z}
\end{eqnarray}
where $I_\pm =({\ell_0 m_\mp - m_0 \ell_\mp})[\omega_\pm(k_0^2 - k_\mp^2)+\ell_\pm N^2({\ell_0}/{\omega_0}-{\ell_\mp}/{\omega_\mp}) ]/({2\omega_\pm k_\pm^2})$. 

Differentiating Equation~(\ref{equation1z}),  one gets
\begin{eqnarray}
{\ddot \Psi_\pm} = I_{+}I_{-} {|}\Psi_0{|}^2 \Psi_\pm - \frac{\nu^2}{4}k_+^2k_-^2 \Psi_\pm- \frac{\nu}{2}(k_+^2+k_-^2){\dot \Psi_\pm}\,. \label{eqfinal} 
\end{eqnarray}
The general solution is $\Psi_{\pm}(t)=A_{1,2}\,\exp{(\sigma t)} +B_{1,2}\, \exp{(\sigma' t)},$  with $\sigma = -{\nu}(k_+^2 +k_-^2)/4  + \sqrt{({\nu}/{4})^2(k_+^2 -k_-^2)^2+I_+I_-|\Psi_0|^2}$ and $\sigma'<0<\sigma$. 

In conclusion, a vanishingly small amplitude noise induces the growth of two secondary waves by a triadic resonant mechanism. Since their sum gives the primary frequency (see Equation~(\ref{temporalcondition})), $\omega_+$ and $\omega_-$ are subharmonic waves. The growth rate of the instability depends on the characteristics of the primary wave, namely its wave vector, its frequency and its amplitude~$\Psi_0$, but also on the viscosity~$\nu$.

\subsubsection{Triads, Resonance Loci and Growth Rates}
\begin{figure}
\begin{center}
\includegraphics[width=\textwidth]{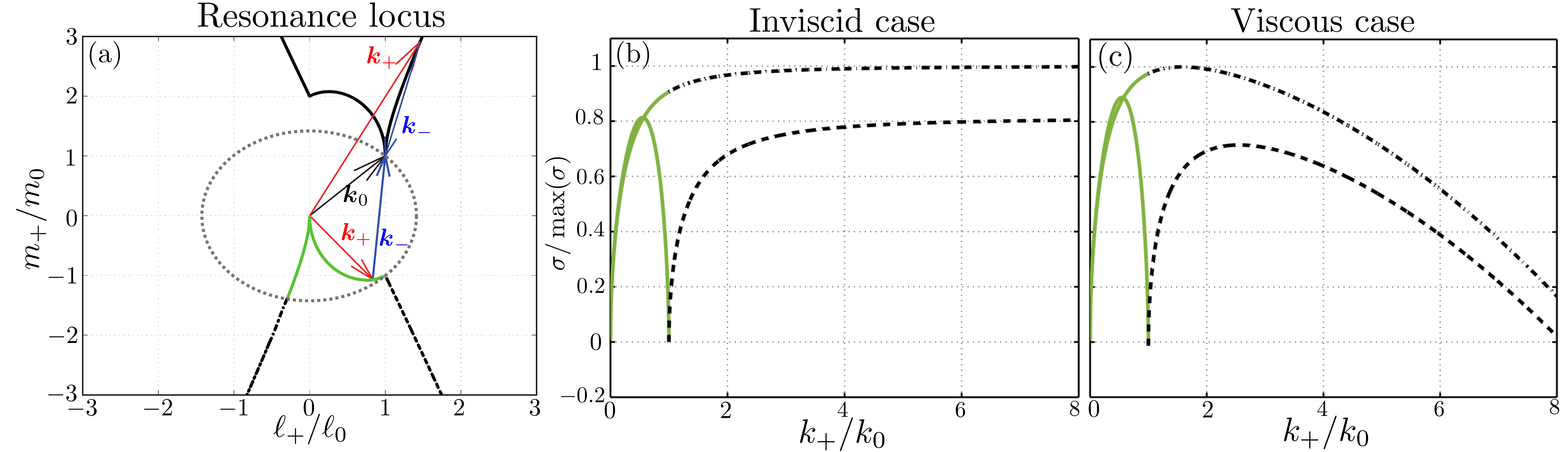}
\vskip -.5truecm
\caption{{(a)  
Resonance locus for the unstable wave vectors $(\ell_+,m_+)$ 
satisfying Equation~(\ref{equation_k1m1}) 
 once the primary wave vector $\mbox{\boldmath $k$}_0=$$(\ell_0,m_0)$ 
  is given. 
Two examples of vector triads ($\mbox{\boldmath $k$}_0$, $\mbox{\boldmath $k$}_+$, $\mbox{\boldmath $k$}_-$) are shown. The dotted curve is defined by $k_+=k_0$. The solid green curves correspond to the central branch, while the dashed and dash-dotted black curves correspond to the external branch.
(b) and (c) Corresponding growth rates $\sigma/\max(\sigma)$ 
 as a function of the normalized wave vector modulus~$k_+/k_0$. 
 (b) presents the inviscid case
 while (c) presents a viscous case corresponding to $\Psi_0/\nu=100$.}
}
\label{dessindelacacouete}
\end{center}
\end{figure}

Using the dispersion relation for internal waves, the temporal resonance condition leads to~\citep{BDJO2013} 
\begin{eqnarray}\label{equfinale}
\frac{ |\ell_0|}{\sqrt{{\ell_0^2+m_0^2}}} & = & 
\frac{{|\ell_+|}}{\sqrt{{\ell_+^2+m_+^2}}} + 
\frac{{|\ell_0 {-}\ell_+|}}{\sqrt{(\ell_0{-}\ell_+)^2+(m_0{-}m_+)^2}}\,,\label{equation_k1m1}
\end{eqnarray}
whose solutions  
are presented in \textbf{Figure~\ref{dessindelacacouete}}a. 
Once the primary wave vector $\mbox{\boldmath $k$}_0$ is defined,
any point of the solid curve corresponds to the tip of the $\mbox{\boldmath $k$}_+$ vector, 
while  $\mbox{\boldmath $k$}_-$ is obtained by   closing the triangle.  
The choice between the labels
 + and - is essentially arbitrary and this leads to the symmetry $\mbox{\boldmath $k$}\rightarrow
\mbox{\boldmath $k$}_0-\mbox{\boldmath $k$}$ in \textbf{Figure~\ref{dessindelacacouete}}{a}. Without loss of generality, we will always call $\mbox{\boldmath $k$}_+$ the largest wavenumber.

One can observe two distinct parts of this resonance locus, characterized by the position of $k_+/k_0$ with respect to  1.
 The wavelength of  the secondary waves generated by the instability can be 
 \begin{itemize}
\item both smaller than the primary wavelength: this case corresponds to the external branch of the resonance locus and
implies an energy transfer towards smaller scales {(represented by black curves in \textbf{Figure~\ref{dessindelacacouete}})}.
 \item one larger and the other one smaller: this case corresponds to the central branch of the resonance locus and implies an energy transfer towards smaller and larger scales {(represented by  solid green curves in \textbf{Figure~\ref{dessindelacacouete}})}.
 \end{itemize} 
 
Among the different possible solutions on the resonance locus, the one expected to be seen experimentally or numerically is the one associated {with} the largest growth rate. In the inviscid case, the most unstable growth rate occurs for $k\rightarrow\infty$, with essentially $\mbox{\boldmath $k$}_+\simeq-\mbox{\boldmath $k$}_-$,
and therefore  $ \omega_+=\omega_-=\omega_0/2$. This ultraviolet catastrophe is healed in the presence of viscosity, which selects a finite wavelength for the maximum growth rate \citep{Hazewinkel2011} 
as shown in Figure~\ref{dessindelacacouete}c.
For typical laboratory scale experiments,  the values of $k_+$ corresponding to significant growth rates are of
the same order of magnitude as the primary wavenumber $k_0$, as can be seen in \textbf{Figure~\ref{dessindelacacouete}c}, with $k_{1}/k_{0}\simeq 1.5$ and  $k_{2}/k_{0}\simeq 2.3$.
Thus, TRI corresponds to a direct energy transfer from the primary wave to small scales where viscous effects come into play, without the need of a turbulent cascade process.

The fact that viscosity has a significant effect on the selection of the excited resonant triad, preventing any large wave number secondary wave to grow from the instability, has been observed  by \cite{BDJO2013} in laboratory experiments on wave beams. However, they also found a different type of triads than those predicted by the previous theoretical arguments. This will be discussed in more detail in the following sections.

\subsubsection{Amplitude Threshold for Plane Wave Solutions}\label{threshold}
The expression for the growth rate $\sigma$ implies that the amplitude of the stream function has to be larger than the critical value 
$|\Psi_c(\ell_+,m_+)|={\nu k_+k_-}/{\sqrt{4I_+I_-}}$ to get a strictly positive growth rate~\citep{Koudella2006,BDJO2013}.
The threshold for the instability is thus given by the global minimum of this function of several variables.
Let us focus on the particular case where $\mbox{\boldmath $k$}_+$ tends to $\mbox{\boldmath $k$}_0$
by considering the following description of the wave vector components 
$\ell_+=\ell_0(1+\mu_0\varepsilon^{\alpha})$ and $ m_+=m_0(1+\varepsilon)$ where $\varepsilon{\ll 1}$, $\alpha\geq 1$, while  $\varepsilon$ and $\mu_0$ are positive quantities. Using the dispersion relation, the temporal and spatial resonance conditions, \cite{BSDBOJ2014} have shown that  $\alpha=2$ is the only acceptable value to balance the lowest order terms. 
Plugging these relations into the expression of $I_\pm$, one gets $I_+=-\ell_0m_0\varepsilon+o(\varepsilon)$ and $I_-=-\ell_0m_0+o(1)$, which leads to $|\Psi_c|=\sqrt{\varepsilon}\, {\nu N}/{(2\omega_0)}+o(\varepsilon^{1/2})$. 
The minimum of this positive expression being zero, it shows that there is no threshold for an infinitely wide wave beam, even when considering a viscous fluid. Plane wave solutions are thus always unstable to this Triadic Resonant Instability.

\subsection{Why does the Finite Width of Internal Waves Beam Matter?}

\begin{figure}
\begin{center}
\includegraphics[width=0.99\textwidth]{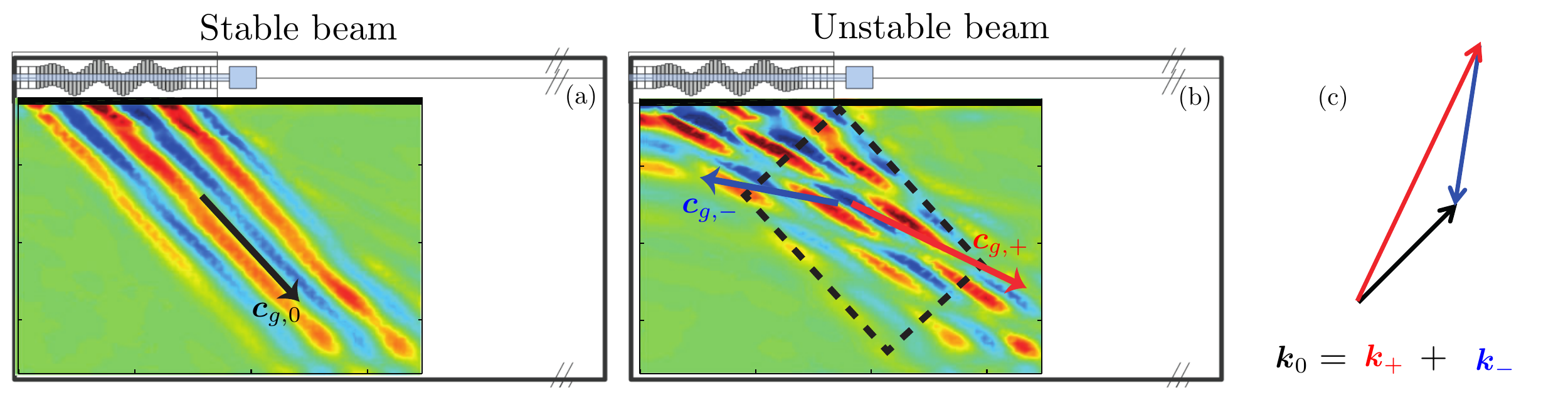}
\vskip -.5truecm
\caption{
Sketch of the experimental set-up showing the wave generator
lying horizontally at the top of the wave tank with a superimposed snapshot of the 
vertical density gradient field. 
(a) The internal wave beam is propagating downward.
(b) The instability of the propagating internal wave beam is 
visible~\citep{BourgetPhD}.
The tilted dashed rectangle corresponds to the control area for
the energy approach of section~\ref{energyapproach}. 
(c) The vector triad with the three arrows representing  
the primary wave vector $\mbox{\boldmath $k$}_0$ (black) and the two 
secondary waves vectors $\mbox{\boldmath $k$}_+$ 
(red) and $\mbox{\boldmath $k$}_-$ (blue).
From this triad, it is possible to deduce the orientation of the 
group velocities of the three different waves as shown in panels (a) and (b).
}
\label{dessindelavitessedegroupe}
\end{center}
\end{figure}

The above theory for the TRI does not take into account the finite width of the experimental beam. Qualitatively, the subharmonic waves can only extract energy from the primary wave if they 
do not leave 
the primary beam before they can extract substantial energy~\citep{BSDBOJ2014}.  The group velocity of the primary wave is aligned with the beam, but the group velocity of the secondary waves is definitely not, and these secondary waves eventually leave the primary wave beam, as illustrated in \textbf{Figure~\ref{dessindelavitessedegroupe}}. This is a direct consequence of the dispersion relation, which relates the direction of propagation to the frequency: a different frequency, smaller  for subharmonic waves, will lead to a shallower angle. 

 Three comments are in order:

i) The angles between primary and secondary waves
strongly influence the interaction time, and thus the instability.

ii) Secondary waves
with small wave vectors, having a larger group velocity~$c_{g,{\pm}}=(N^2-\omega_{\pm}^2)^{1/2}/k_{\pm}$,  leave the primary wave beam more rapidly
and have less time to grow in amplitude. 
Such solutions will therefore {be} less likely to develop, {opening the door to} stabilization of the primary wave by {the} finite width effect.
This clarifies why experiments with the most unstable secondary waves on the internal branch (small wave vector case) of the resonance locus (\textbf{Figure~\ref{dessindelavitessedegroupe}}) were found to be stable~\citep{BDJO2013} contrary to the prediction for plane waves.  This decisive role of the group velocity of the short-scale subharmonic waves was identified long ago by \cite{McEwanPlumb1977}. 

iii) At the other end of the spectrum, small wavelengths are more affected by dissipation and will also be less likely to be produced by TRI. Consequently, only a window of secondary wavelengths is possibly produced by TRI.

\subsection{Energy Approach}\label{energyapproach}

A simple energy balance proposed by~\cite{BSDBOJ2014} makes possible an insightful and more quantitative estimate for the most unstable triad.
We introduce the tilted rectangle shown in \textbf{Figure~\ref{dessindelavitessedegroupe}} as control area 
{(denoting $W$ the perpendicular beam width)} and we neglect the spatial attenuation of the primary wave in this region (``pump-wave'' approximation). 
Since secondary waves do not propagate parallel to the primary beam,  they exit the control area from the lateral boundaries without compensation.
Equation (\ref{equation1z}) is  thus modified as follows
\begin{eqnarray}
\frac{{\rm d}\Psi_\pm}{{\rm d}t} & =&|I_\pm|\Psi_0\Psi_\mp^*-\frac{\nu}{2} k_\pm^2\Psi_\pm-\frac{|{\mbox{\boldmath $c$}_{g,\pm}}\cdot{\mbox{\boldmath $e$}_{k_0}}|}{2W}\Psi_\pm \ . \label{equation3}
\end{eqnarray}
The first term represents the interaction with the other plane waves of the triadic resonance, the second term is due to viscous damping while the third one accounts for the energy leaving the control area.

 One finds here also exponentially growing solutions with a positive growth rate slightly modified as 
$\sigma
^*=-\left(\Sigma_++\Sigma_-\right)/4
+\sqrt{\left(\Sigma_+-\Sigma_-\right)^2/16+|I_+||I_-||\Psi_0|^2}, $ 
 in which the effective viscous term now reads $\Sigma_\pm=\nu k_\pm^2+
 {|{\mbox{\boldmath $c$}_{g,\pm}}\cdot{\mbox{\boldmath $e$}_{k_0}}|}/{W}$. 
 The finite width of the beam  is responsible for a new term
 characterizing the transport of the secondary waves energy out of the interaction region. For infinitely wide wave beams ($W\rightarrow+\infty$), one recovers the growth rate $\sigma$ obtained in the plane wave case. In contrast, when the beam becomes narrow ($W\rightarrow0$), the growth rate decreases to zero, leading to a stabilization.
 
The finite width of a wave beam increases therefore its stability, owing to the transport of the secondary waves out of the triadic interaction zone of the primary wave beam before they can extract substantial energy. This interaction time scales directly with the perpendicular beam width, $W$ as can be seen from the expression of $\sigma^*$.
 
\subsection{Theory in the Nearly Inviscid Limit}\label{TheoryintheNearlyInviscidLimit}

A  beautiful weakly nonlinear asymptotic analysis of the finite width effect on TRI
has been recently proposed by \cite{Karimi2014}. 
Mostly interested by oceanic applications, they look for subharmonic perturbations in the form of fine-scale, nearly monochromatic 
wavepackets with frequency close to one half of the primary frequency; in this limit, usually
called Parametric Subharmonic Instability (see the sidebar distinguishing TRI vs. PSI), $\omega_\pm\simeq \omega_0/2={N}(\sin\theta)/2={N}\sin\phi$, that defines the angle~$\phi$ with the vertical of the wave vectors $\mbox{\boldmath $k$}_\pm$ of opposite directions.

The key ingredient in the analytical derivation is to take advantage of the scale difference between the width of the primary beam~$W$ and the very small carrier wavelength of the subharmonic wave packets
$\lambda_\pm=2\pi/k_\pm$. The small amplitude expansion is thus characterized by the small
parameter $\mu=\lambda_\pm/(2\pi W)$.
They consider an expansion with not only the underlying wave beam but also the superimposed subharmonic wavepackets that appear to order $\mu$. 
The derivation of the wave-interaction equations leads to six coupled equations for 
the two primary beam envelopes (two because of the two phases)
and the four ($2\times2$) subharmonic wavepacket envelopes. Fortunately, this system can be reduced at leading order 
to only three coupled equations with three unknowns.

Taking a distinguished limit in which not only the amplitude of the primary wave, but also the nonlinear, dispersive and viscous terms are all small, they obtain a reduced description of the dynamics.
The strategy is as usual to choose the scaling in order to get comparable magnitudes of the different terms.
Interestingly, as only the quadratic interaction is potentially destabilizing for the primary beam, they compare 
it with the advection term for subharmonic waves: the former being smaller, this confirms
 that the resonant interaction cannot feed the instability in the limited time during which perturbations are in contact with the underlying beam.
Beams with a general profile of finite width are thus stable to TRI.

Next, they consider the case of beams with profiles in the form of a monochromatic carrier with $O(1)$ wavelength, modulated by a confined envelope. Functions of the cross-beam direction $\eta$ (see~\textbf{Figure~\ref{profilselonetabb}}) and of the appropriate slow time $\tau$, the complex envelopes  $\Psi_0(\eta,\tau)$ and $\Psi_\pm(\eta,\tau)$, of
the primary and secondary waves are thus generalizations of the plane wave solutions considered in section~\ref{derivEquPlaneWaves}. We recover these solutions with an envelope function independent of the cross-beam coordinate $\eta$, while an internal wave beam (such as the one in \textbf{Figure~\ref{dessindelavitessedegroupe})} will correspond to $\Psi_0(\eta)=1/2$ for $|\eta|<1/2$ and zero otherwise. Introducing the appropriate change of variables
and rescaling of the relevant variables,
the beam envelope $\Psi_0$ and the complex subharmonic envelopes $\Psi_\pm$ are linked through the following three coupled {dimensionless} equations
\begin{eqnarray}
\frac{\partial \Psi_\pm}{\partial\tau}&=&-\mbox{\boldmath $c$}_{g,\pm}\cdot\mbox{\boldmath $e$}_{k_0}\frac{\partial \Psi_\pm}{\partial\eta}-\overline{\nu} \kappa^2 \Psi_\pm
+i\frac{{N}\kappa^2}{\omega_0}\sin^2\chi \, |\Psi_0|^2\Psi_\pm+\varpi \Psi_0\Psi^*_\mp,\label{equationfora}\\
\frac{\partial \Psi_0}{\partial\tau}&=&-2\varpi \Psi_+ \Psi_-, \label{equationforq}
\end{eqnarray}
where $\overline{\nu}$ is the renormalized viscous dissipation and $\kappa=2/\mu$
a rescaled wavenumber modulus. $\chi=\theta-\phi$ and $\varpi=\sin\chi 
\cos^2\left({\chi}{/2}\right)$ 
 are two geometrical parameters, while  $\tau$  is the appropriate slow time for the evolution of the subharmonic wave packets and $\eta$ the appropriate cross-beam coordinate.

In the appropriate distinguished limit identified by~\cite{Karimi2014}, 
the nonlinear term is balanced as in section~\ref{energyapproach} by the viscous term,
but also by the transport term.
The coupling between the evolution equations occurs through the nonlinear terms, which allow energy
exchange between the underlying beam and subharmonic perturbations.
The subsequent behavior of the complex envelopes~$\Psi_\pm$ determines the stability of the beam: if they are able
to extract energy via nonlinear interaction with~$\Psi_0$ at a rate exceeding the speed of
linear transport and viscous decay, the beam is unstable.

From this system, it is in principle possible to study the stability of any profile:

\begin{enumerate}

\item For example, a time independent beam $(\Psi_0(\eta),\Psi_\pm=0)$ is a steady state solution of this system of three equations.
The study of its stability relies on looking for the normal mode solutions $\Psi_\pm\propto\exp(\sigma\tau)$.
For a plane wave,  one obtains the growth rate $\sigma=\sin\chi\cos^2\left({\chi}/{2}\right)/2-\overline{\nu} \kappa^2$.  A subtle point was carefully emphasized by~\cite{Karimi2014}. 
The above expression of $\sigma$ seems independent of the wave vector disturbance 
$\kappa$ {in the inviscid limit, but the derivation has} extensively 
 used the hypothesis of fine-scale disturbances, that will of course break down for $\kappa\ll1$.
The maximum growth rate is indeed attained for finite but small $\kappa$. 
Uniform beams (internal plane waves
with a confined envelope) are unstable if the beam is wide enough.

\item \cite{Karimi2014} provide also the solution of
the initial value problem for a beam 
with $\Psi_0(\eta)$ tending towards zero as $\eta$ tends to infinity.
They show the existence of a minimum value for the unstable perturbation wavenumber $\kappa_{\mbox{\tiny min}}=\pi c_{g,\pm}/(2\varpi W\int_{-\infty}^{+\infty}\Psi_0(\eta)\mbox{d}\eta)$, corresponding to a maximum wavelength. Therefore, the possible spatial scale window for secondary wavelengths shrinks towards smaller scales as the beam is made narrower.
Outside this range, no instability is possible even in the inviscid case.

\item They derive also analytically the minimum width explicitly for the top-hat profile
used in the experiments by~\cite{BDJO2013}  ($\Psi_0(\eta)=1/2$ for $|\eta|<1/2$ and zero otherwise as
shown in \textbf{Figure~\ref{dessindelavitessedegroupe}}). They argue that the existence of 
a minimum width
is valid for a general profile. This minimum is dependent on the beam shape.

\end{enumerate}

To summarize, internal plane waves
with a confined envelope are unstable if the beam is wide enough,
while weakly nonlinear beams with a general but confined profile (i.e. without any dominant carrier wavenumber) are stable to short-scale subharmonic 
waves.

\subsection{Effect of a mean advective flow} 
\cite{LerissonChomaz2017} have recently studied theoretically and numerically the Triadic Resonant Instability of an internal gravity  beam in presence of a mean advective flow.
They keep constant the local wave vector and wave frequency in the frame moving with the fluid 
in order to encompass both tidal flows and lee waves. 

Their main result is that, by impacting the group velocity of the primary and secondary waves, 
the mean advection velocity modifies the most unstable triads. They have
predicted and confirmed numerically that a strong enough advective flow 
enhances the instability of the central branch (leading to large scale mode since one secondary wavelength
is larger than the primary one) with respect
to the  external branch.
However, the model is not able to explain the existence of an interesting stable region, at intermediate velocity in their numerical simulations.
To go beyond, it would be necessary to take into account the spatial growth of the secondary waves within the
internal wave beam. Such a theory relying
on the extension of the classical absolute or convective instability is still to be derived.

\subsection{Effect of the rotation}
\subsubsection{Theoretical study}

When one includes Coriolis effects due to Earth's rotation at a rate~$\Omega_C$, the dispersion relation of internal waves is modified
and this has of course consequences on the group velocity, which we showed to be intimately associated to the stability of internal wave beams.
\cite{BordesFast2012} have reported experimental signatures of Triadic Resonant Instability
of inertial gravity beams in homogenous rotating fluid. It is thus expected that TRI
will also show up when considering stratified rotating fluid.

Assuming invariance in the transverse $y$ direction, the flow field may be written as  $(u_x,u_y,u_z) = ( \partial_z \psi,u_y,-\partial_x\psi )$ with $\psi(x,z,t)$ the streamfunction of the non-divergent flow in the vertical plane, and $u_y(x,z,t)$ the transverse velocity.
Introducing the  Coriolis parameter $f=2 \Omega_C\sin\beta$, where $\beta$ is
the latitude, the dynamics of the flow field is given by the following system of three equations
\begin{eqnarray}
\partial_t b + J(b,\psi) - N^2 \partial_x \psi &=& 0, \label{eq_gravity_1rot}\\
\partial_{t}\nabla^{2} \psi + J(\nabla^{2} \psi , \psi) -f\partial_z u_y&=& - \partial_x b+\nu \nabla^{4} \psi,\label{equationenpsirot} \\
 \partial_t u_y+ J(u_y,\psi) +f\partial_z \psi &=&\nu \nabla^{2} u_y. \label{equationenrhorot}
\end{eqnarray}
in which the equation for the buoyancy perturbation is not modified, 
while Equation~(\ref{equationenpsi}) has been modified and is now coupled to the dynamics of
the transverse velocity $u_y$.

As previously, one can study beams of general spatial profile, corresponding to 
the superposition of time-harmonic plane waves with a dispersion relation~(\ref{eq_disp_gravity_theta}) modified
into  $ \omega^2= N^2\sin^2\theta+f^2\cos^2\theta.$ 
The next step is again to look for subharmonic perturbations in the form of fine-scale (with respect to the width of the beam), nearly monochromatic 
wavepackets with frequency close to half the primary frequency. It is straightforward to see that subharmonic waves propagate with an inclination~$\phi$ given
by $\sin\phi=((\omega_0^2/4-f^2)/(N^2-f^2))^{1/2}$ that vanishes when $\omega_0/2= f$, i.e. at the critical latitude $\beta\simeq28.8^\circ$~\citep{MacKinnonWinters2005}.
The modulus of the group velocity of subharmonic waves 
$c_{g,\pm}=(N^2-f^2) \sin(2\phi)/(\omega_0 k_\pm)$ will thus also vanish at this latitude.
The rotation reducing dramatically the ability of subharmonic waves to escape, it may seriously
reinforce the instability. 

\cite{Karimi2017} have shown that it is possible to reproduce the asymptotic analysis of the Triadic Resonant Instability with the inclusion
of Earth's rotation (see also \cite{KarimiPhD2015}). 
One ends up with an unchanged equation~(\ref{equationforq}) for the dynamics of the primary wave,
while the coupled dynamics of subharmonic waves is modified as follows
\begin{eqnarray}
\frac{\partial \Psi_\pm}{\partial\tau}=
-\mbox{\boldmath $c$}_{g,\pm}\cdot\mbox{\boldmath $e$}_{k_0}
\frac{\partial\Psi_\pm}{\partial\eta}
+\frac{i}{2}\frac{3f}{\kappa^2{N}}\frac{{\partial^2} \Psi_\pm}{\partial\eta^2}-\overline{\nu}\kappa^2 \Psi_\pm
+i\delta{\kappa^2}{} \, \left|\frac{\partial \Psi_0}{{\partial\eta}}   \right|^2\psi_\pm-\gamma 
\frac{\partial^2\Psi_0}{{\partial\eta^2}}\Psi^*_\mp,\label{equationforarot}
\end{eqnarray}
with 
 $\delta$ and $\gamma$ two parameters depending on the Coriolis parameter $f$, which vanish when $f$ tends to zero.
The important modification is the appearance, on the right-hand side, of the second linear term due to dispersion. 
It is important here because the first one may disappear since
 the projection of the respective group velocity $\textbf c_{g,\pm}$ of subharmonic envelopes $\Psi_\pm$ on
the across beam direction $\textbf e_{k_0}$ may vanish.

\cite{Karimi2017} consider first weakly nonlinear sinusoidal wavetrains, emphasizing
two interesting limits: the 
case far from the critical latitude allows one to recover
the results of section~\ref{TheoryintheNearlyInviscidLimit}
in which there is no preferred wavelength of instability in the inviscid limit. 
On the other hand, when the group velocity of perturbations vanishes at the critical 
latitude, energy transport is due solely to second-order dispersion. This process of energy transport leads to the selection
of a preferred wavenumber, independent of damping effects,
which may suppress the instability for a sufficiently large damping factor, 
permitting the underlying wave to survive the instability. They obtained an expression for
the growth rate identical to the result in the inviscid limit of~\cite{YoungTsangBalmforth2008}. It explains that, at the critical latitude, additional physical factors, such as scale-selective dissipation, must become important. This result has been shown numerically by~\cite{Hazewinkel2011} in agreement with in situ measurements~\citep{Alford2007}.

For beams, there is always a competition between energy extraction 
from the beam, which varies with beam profile, and the proximity to the critical latitude,
without forgetting the viscous effects on the fine-scale structure of disturbances.
Relying on numerical computations, it is possible to predict the stability properties
for a given profile.
In general, it turns out that rotation plays a significant role in dictating energy transfer from an internal wave to fine-scale disturbances via TRI under resonant configurations. 

\subsubsection{Experimental  study}

Until now, the only laboratory experiment that studied the influence of rotation on the triadic instability of inertia-gravity waves, in a rotating stratified fluid, was performed recently by~\cite{Maureretal2016}. In this study, the   set-up that was used by~\cite{BDJO2013,BSDBOJ2014} was placed on a rotating platform, with a range of rotation rates from 0 to 2.16 rpm, allowing the dimensionless Coriolis parameter, $f/N$, to vary in a range from 0 to 0.45.

One of their main findings is the observation that the TRI  threshold in frequency is lowered (by about 20\%), compared to the non-rotating case. An extension of the energy approach developed in section~\ref{energyapproach}
 to the rotating case confirms this observation, by showing that the finite-size effect of the beam width is reduced when rotation increases. This enhancement of TRI only applies to a limited range of rotation rate since, when the rotation rate overcomes half the primary wave frequency, TRI is forbidden due to the dispersion relation not allowing the existence of the lowest-frequency secondary wave. The competition between this limit and the finite-size effect reduction by rotation results in a minimum value for the frequency threshold. The position of this minimum, observed around $f/\omega_{0}\simeq 0.35$ in the experiment, depends on the Reynolds number, defined as $Re=\Psi_{0}/\nu$. The transposition of this result to high Reynolds number situations like in the ocean shows that the TRI enhancement is then localized in a narrow Coriolis parameter range, with $f=\omega_{0}/2$, thus recovering the critical latitude phenomenon.

When global rotation is applied to the fluid, another interesting feature is that it creates an amplitude threshold for TRI. Indeed, it was discussed in section~\ref{threshold} that in the absence of rotation, plane waves were always unstable. However, as shown in this section, the instability at very low amplitude occurs when  $\mbox{\boldmath $k$}_{+}$  tends towards $\mbox{\boldmath $k$}_{0}$, which implies $\omega_{+}\rightarrow \omega_{0}$ and therefore $\omega_{-}\rightarrow 0$. But when there is rotation in the system, a zero-frequency subharmonic wave is no longer allowed, hence the appearance of a threshold in amplitude, which increases with $f/N$.

\section{STREAMING INSTABILITY}\label{StreamingInstability}

\subsection{Introduction}
Another important mechanism for the instability of internal gravity waves beams is the generation of a mean flow,
also called streaming instability\footnote{This should not be mixed-up with the mechanism for planetesimal formation in astrophysics.} by \cite{KataokaAkylas2015}. 
\cite{LighthillBook} and \cite{andrews1978} noticed in early times that 
 internal gravity wave beams share several properties with acoustic wave beams.
In particular, both kinds of waves may be subject to streaming in the presence of  dissipative effects. Streaming refers here to the
emergence of a slowly evolving, non-oscillating, Eulerian flow forced by nonlinear interactions of the oscillating wave-beam with itself~\citep{nyborg1965acoustic,lighthill1978}.  As reviewed in \cite{riley2001}, it is now recognized that streaming occurs actually in a variety of flow models; it remains an active field of research for both theoretical~\citep{xie2014boundary} and experimental~\citep{Squires2013,moudjed2014scaling,Wunenburger} points of view. 

The fact that dissipative effects are required to generate irreversibly
a mean flow through the nonlinear interactions of a wave beam with
itself can be thought of as a direct consequence of ``non-acceleration''
arguments that came up in the geophysical fluid dynamics context fifty
years ago with important contributions from~\cite{charney1961propagation,eliassen1961transfer,andrews1976}, among others. 
 \cite{plumb1977interaction} used those ideas to propose
an idealized model for the quasi-biennal oscillation (QBO), together
with an experimental simulation of the phenomenon \citep{PlumbMcEwan1978}.
 The oscillations require more than one wave beam, but \cite{plumb1977interaction} discussed
first how a single wave beam propagating in a vertical plane could
generate a mean flow. He predicted the vertical shape of this mean
flow, emphasizing the important role played by the wave attenuation
through dissipative effects. The
experiment by \cite{PlumbMcEwan1978} may be thought of 
as the first quantitative observation of streaming in stratified fluids.

Those examples correspond, however, to a very peculiar instance of streaming,
with no production of vertical vorticity. By contrast, most applications
of acoustic streaming since the earlier works of \cite{eckart1948vortices} and \cite{westervelt1953theory}
involve the production of vorticity by an irrotational
wave. 
As far as vortical flows are concerned, \cite{LighthillBook} noticed important analogies between acoustic waves and internal gravity waves:  in both cases, vortical flows and propagating waves are decoupled at a linear level {in the inviscid limit}, and steady streaming results from viscous attenuation {of the wave amplitude. In particular, \cite{LighthillBook} noticed that streaming could generate a flow with vertical vorticity}. However, experimental observation of the emergence of a vortical flow in stratified fluids through this mechanism remained elusive until recently.
While studying the internal wave generation process via a
tidal flow over 3D seamounts in a stratified fluid, \cite{ZhankKingSwinney2007} 
observed a strong flow in the plane perpendicular to
the oscillating tidal flow.  For low forcing, this flow was found to
be proportional to the square of the forcing amplitude. That led them
to invoke nonlinear interactions, either between the internal wave beam and itself, or between internal waves and the viscous boundary layer.
The analysis was not pursued further, and the sign of the vorticity generated, opposite to the one discussed in next subsections, remains puzzling. 

A few years later, studying the reflection of an internal wave beam on a sloping bottom, Grisouard
and his collaborators have also discovered this mean-flow generation
in experiments \citep{GrisouardPhD,Leclairetal2011,Grisouardetal2013}. 
The basic configuration was an uniform beam
reflecting onto a simple slope in a uniformly stratified fluid. As
predicted \citep{DauxoisYoung1999,Gostiaux2006}, the interaction
between the incident and reflected waves produced harmonic waves,
thereby reducing the amplitude of the reflected wave. However, more
surprisingly, they found that the reflected wave was nearly absent
because a wave-induced mean flow appeared in the superposition region
of the incident and reflected waves, progressively growing in amplitude.
Comparing two- and three-dimensional numerical simulations, they showed
that this mean flow is of dissipative origin\footnote{
Note however that transient mean flows can be generated by inviscid motion in the wake of a propagating internal wave packet~\citep{Bretherton1969,vandenBremerSutherland2014}.} and three-dimensional.
Its presence totally modifies the two-dimensional view considered
in the literature for reflection of internal waves. Indeed, there
has been many interesting theoretical studies of internal gravity
waves-mean flow interactions \citep{Bretherton1969,LelongRiley1991,Akylas2003}, but none of them considered the effect of dissipation in three dimensions.

The complete and theoretical understanding of the generation of a
slowly evolving vortical flow by an internal gravity wave beam was
possible using an even simpler set-up that we describe in the following
section. \cite{Bordes2012} reported observations of a strong mean
flow accompanying a time-harmonic internal gravity beam, freely propagating
in a tank significantly wider than the beam. We describe below in detail the experimental
set-up and the observations, together with two related theories by \cite{Bordes2012} and \cite{KataokaAkylas2015}, which describe well the experimental results, by providing the spatial structure and temporal evolution of the mean flow and illuminating the mechanism of instability. Those approaches bear strong similarities with the result obtained
by \cite{GrisouardBuhler2012}, who used generalized Lagrangian mean theory, in order to describe the emergence of a vortical flow in the presence of an oscillating flow of a barotropic tide above topography variations.



\subsection{Experimental Observations}

\cite{Bordes2012} have studied an internal gravity wave beam 
of limited lateral extent propagating along a significantly wider stratified fluid tank.
Previously,  most experimental studies that were using the same internal wave generator~\citep{Gostiaux2007,Mercieretal2010} were quasi-two-dimensional (beam and tank of equal width) and therefore
without significant transversal variations.

\begin{figure}
\begin{center}
\includegraphics[width=\textwidth]{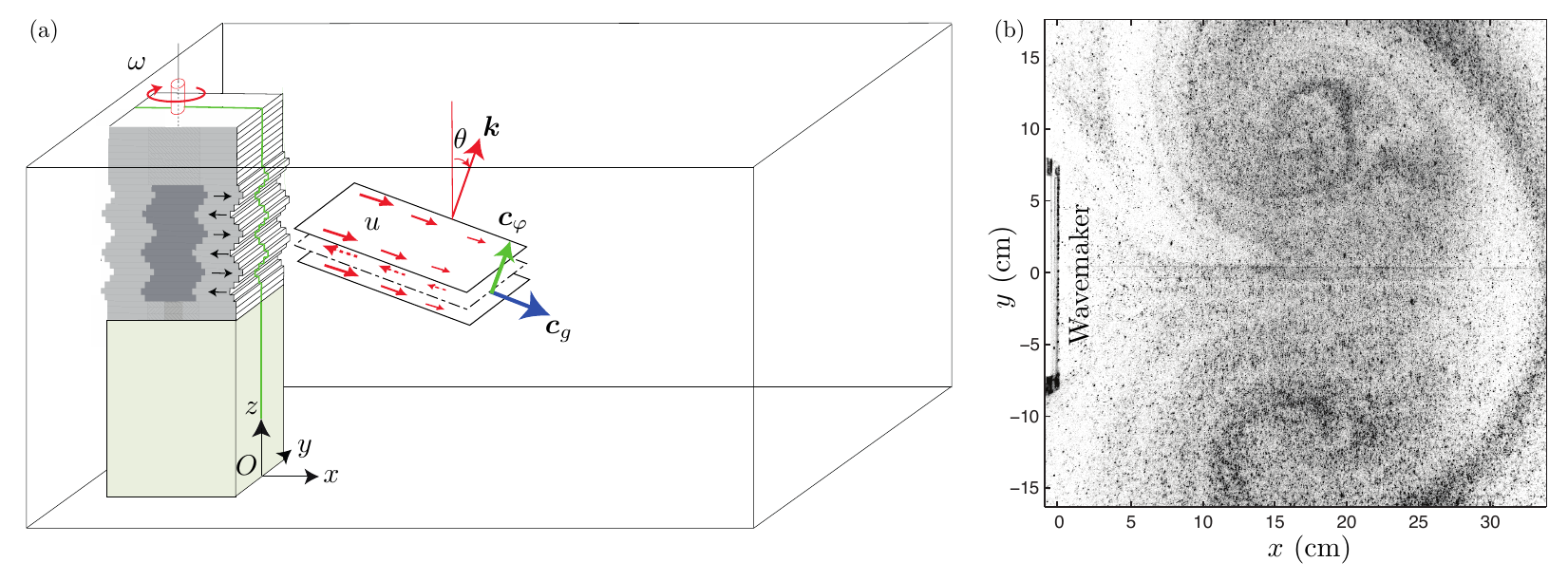}
\caption{(a) Schematic representation of the experimental set-up with the generator on the left of the tank. 
(b) Top view of the particle flow in a horizontal plane at intermediate depth.
}
\label{SetupmanipGuilhem}
\end{center}
\end{figure}

 \textbf{Figure~\ref{SetupmanipGuilhem}a} presents a schematic view of the experimental set-up in which one can see the generator, the tank and the representation of the internal wave beam
generated (see~\cite{BordesPhD} for additional details).
The direct inspection of the flow field shows an unexpected and spontaneously generated pair of vortices, 
emphasized in \textbf{Figure~\ref{SetupmanipGuilhem}b} by the tracer particles dispersed in the tank to visualize the flow field using particle image velocimetry. 
This structure is actually a consequence of the generation of a strong mean flow. This experiment provides therefore
an excellent set-up to carefully study the mean-flow generation and to propose a theoretical understanding that 
explains the salient features of the experimental observations.

\begin{figure}
\begin{center}
\includegraphics[width=\textwidth]{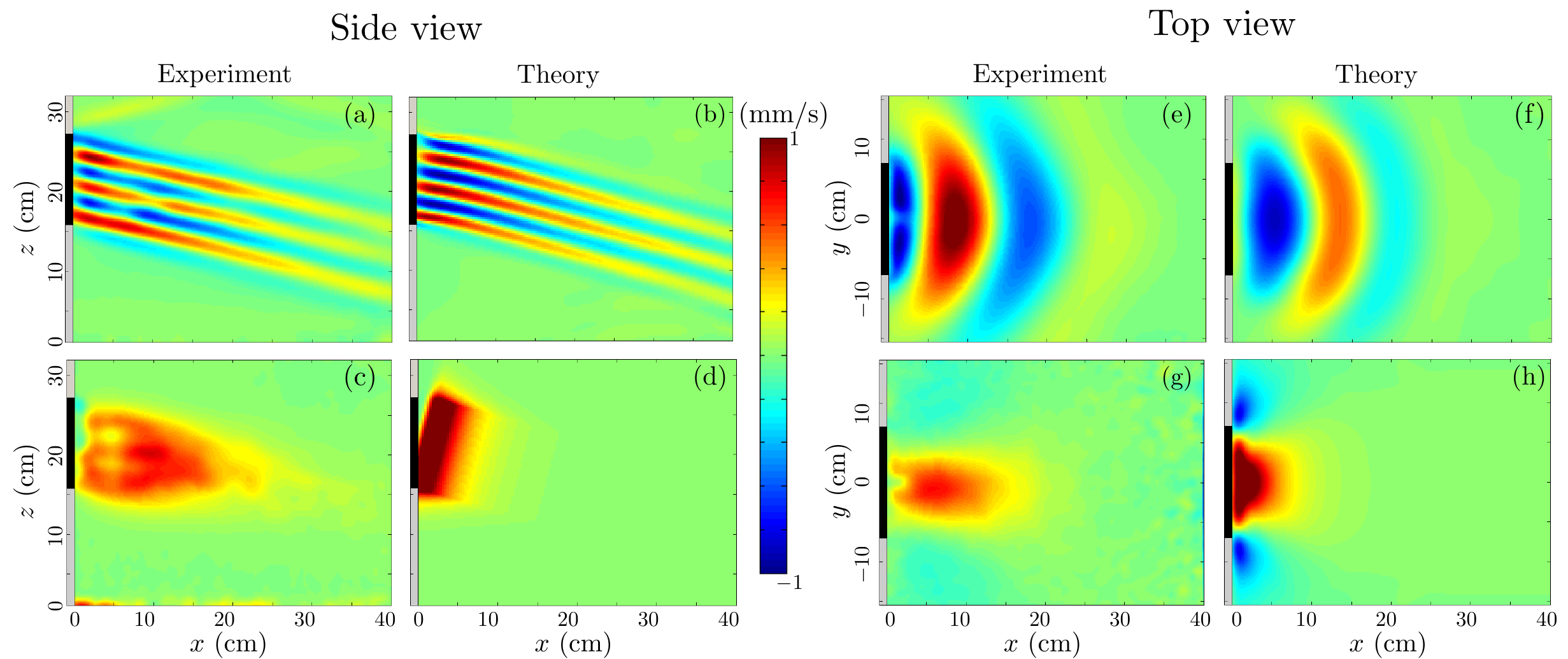}
\caption{Experimental (a,c,e,g) and theoretical (b,d,f,h) horizontal velocity fields $u_x$ for the primary wave (top plots, obtained by filtering \citep{Mercieretal2008} the velocity field at the forcing frequency) and the mean flow 
(bottom plots, obtained by low-pass filtering the velocity field) as reported respectively in \cite{Bordes2012} and  \cite{KataokaAkylas2015}. The four left panels present the side view, while the right ones show the top view. The wave generator is represented in grey with its moving part in black.} 
\label{ResultmanipGuilhem}
\end{center}
\end{figure}

These observations are summarized in \textbf{Figure~\ref{ResultmanipGuilhem}}, which shows side and top views 
not only of the generated internal wave beam  but also of its associated mean flow. One sees that the wave part of 
the flow is monochromatic, propagating at an angle $\theta$ 
 and with an amplitude 
varying slowly in space compared to the wavelength~$\lambda$.
 These waves are accompanied by  a mean flow with a  jet-like structure, {in the direction of the horizontal propagation of waves,}  together with a weak recirculation outside the wave beam. Initially produced inside the wave beam, this dipolar structure corresponds to the  spontaneously generated vortex shown in \textbf{Figure~\ref{SetupmanipGuilhem}b}.  Moreover, the feedback of the mean-flow on the wave leads to
 a transverse bending of wave beam crests that is apparent in \textbf{Figures~\ref{ResultmanipGuilhem}{e}} and~\textbf{\ref{ResultmanipGuilhem}{f}}.

\subsection{Analytical Descriptions}

\subsubsection{A preliminary multiple scale analysis}

Taking advantage of the physical insights provided by the experiments, \cite{Bordes2012} have proposed 
an approximate description that uses a time-harmonic wave flow with a slowly varying amplitude in space.
The problem contains two key non-dimensional numbers, the Froude number $U/\lambda N$ and  the ratio $\nu/\lambda^2 N$ between the wavelength  $\lambda$ and the attenuation length scale of the wave beam due to viscosity, $\lambda^{3}N/\nu$~\citep{Mercieretal2008}. For analytical convenience, they considered a distinguished limit with the small parameter $\varepsilon=Fr^{1/3}$, 
together with the scaling $\nu/\lambda^2 N=\varepsilon /\lambda_\nu$ where $\lambda_\nu\sim 1$. As usual, the appropriate scaling in the small parameter~$\varepsilon$ is 
deduced from a mix of physical intuition and analytical handling of the calculations. In their case, they were looking for a regime with small nonlinearity and  dissipation.
In terms of the velocity components $(u_x,u_y,u_z)$, the buoyancy $b$ and the vertical vorticity $\Omega=\partial_x u_y -\partial_yu_x$,
the governing dimensionless equations in this three-dimensional setting read now
\begin{equation}
\nabla_{H}\cdot\mbox{\boldmath $u$}_{H}=-\partial_z u_{z},
\label{eq:divAD}
\end{equation}
\begin{equation}
\partial_{t} b+\varepsilon^{3}\left(\mbox{\boldmath $u$}\cdot\nabla b\right)+u_z=0,
\label{eq:BAD}
\end{equation}
\begin{equation}
\partial_{t}\Omega+\varepsilon^{3}\left(\mbox{\boldmath $u$}_{H}\cdot\nabla_{H}\Omega+\left(\nabla_{H}\cdot\mbox{\boldmath $u$}_{H}\right)\Omega+\partial_{x}\left(u_z \partial_z u_{y} \right)-\partial_{y}\left(u_z \partial_z u_x\right)\right)=\varepsilon\lambda_{\nu}^{-1}\nabla^2 \Omega,\label{eq:OmegaAD-1}\end{equation}
\begin{equation}
\nabla^2\partial_{tt} u_z+\nabla^2_{H}u_z=\varepsilon\lambda_{\nu}^{-1}\nabla^4 \partial_tu_{z}-\varepsilon^{3}\left(\partial_{t}\left(\nabla^2_{H}\left(\mbox{\boldmath $u$}\cdot\nabla {u_z}\right)-\partial_{z}\nabla_{H}\left(\mbox{\boldmath $u$}\cdot\nabla\mbox{\boldmath $u$}_{H}\right)\right)+\nabla^2_{H}\left(\mbox{\boldmath $u$}\cdot\nabla b\right)\right),
\label{eq:WAD}
\end{equation}
in which the index $H$ in $\mbox{\boldmath $u$}_{H}$ and $\nabla_{H}$  reduces to the horizontal velocity field, gradient or Laplacian operator.   

Introducing rescaled spatial and time coordinates, a multiple scale analysis 
is now at hand. 
Looking for a flow field in perturbation series
$u_r=u_r^{0}+\varepsilon u_r^{1}+o(\varepsilon)$ for $r=x$, $y$ or $z$ with 
a priori $u_y^{0}=0$ as suggested by the structure of the beam,
together with the vertical vorticity field 
$\Omega=\varepsilon^{2}\Omega_{2}
+\varepsilon^{4}\Omega_{4}
+o(\varepsilon^{4}),$  a tedious but straightforward application of the multiple scale framework (with $x_i=\varepsilon ^i x$ and $t_i=\varepsilon ^i t$)
 gives then, to the first three orders, the structure of the beam:
the first order $\varepsilon^0$ provides the  expressions for $u_x^0$ and $u_z^0$,
the second order $\varepsilon^1$ gives the expression for $u_y^1$ and finally
order $\varepsilon^2$ shows that $u_z^0$ does not depend on the slow timescale $t_2$.

Nonlinear terms contribute a priori for the first time to order $\varepsilon^3$, but one interestingly finds again (see section~\ref{NLterms})
that  they  vanish to this order. 
To order $\varepsilon^4$, one obtains 
that the term independent of the slow time $t_0$ vanishes and, thus, nonlinear terms do not induce a mean flow to this order either.
It is only to order~$\varepsilon^5$ that nonlinear terms directly contribute to the mean-flow generation. 
 The governing equation 
of the vortical flow induced by the mean flow is then given in the original dimensional units by
\begin{equation}
\partial_t \overline\Omega =\frac{\partial_{xy}\,{\cal U}^2}{(2\cos\theta)^2}+\nu \nabla^2\overline\Omega\label{equationmeanflowGuilhem}
\end{equation}
where the overline stands for the filtering over one period and ${\cal U}(x,y)$ is the 
amplitude of the wave envelope.

Several conclusions can be directly inferred from this analysis:

i) As emphasized by the first term on the right-hand side, nonlinear terms are crucial as a source of vertical vorticity. 
 Note that one recovers that 
 the amplitude of the mean flow is proportional to the square of the wave amplitude as has been 
 invoked from experimental
\citep{ZhankKingSwinney2007} 
 or theoretical \citep{BuhlerBook} results.

ii) The variations of the wave field in the $y$-direction (implying $\partial_y\neq0$) are 
necessary for nonlinearities to be a source of vertical vorticity. This illuminates why three-dimensional effects
are crucial and therefore why no mean-flow generation was noticed in two dimensions~\citep{Mercieretal2010,Grisouardetal2013}.

iii) Finally, the viscous attenuation of the wave field in the $x$-direction (implying $\partial_x\neq0$) is also necessary to produce vertical vorticity. In actual experiments, variations of the amplitude in the $x$-direction can also come from finite size effects, but are not sufficient to generate a mean flow.

One drawback of the above approach, however, is that it does not describe the feedback of the mean flow
on the waves. For this reason, the approach becomes inconsistent at long
time in the far field region. 

The above combined experimental and analytical proof of the key role played by viscous attenuation and lateral variation 
of the wave beam amplitude in the generation of the observed mean flow has therefore 
motivated 
a more careful asymptotic
expansion  by~\cite{KataokaAkylas2015},
taking into account the two-way coupling between waves and mean-flow.
This two-way coupling accounts for the horizontal bending of the wave
mean-field in \cite{Bordes2012} experiments, as explained in section \ref{completemodel}. 

\subsubsection{Stability to three-dimensional modulations}\label{Kataoka2013}

Initially, \cite{KataokaAkylas2013} were  interested in three dimensional
perturbations of internal wave beams. Specifically, they studied the
stability of uniform beams subject to oblique modulations which vary slowly in the along-beam
and the horizontal transverse directions. Results turned out to be fundamentally different from that 
of purely longitudinal modulations considered in~\cite{Akylas2003}. Because of the 
presence of transverse variations, a resonant interaction becomes 
possible between 
the primary beam and three dimensional
perturbations. 
Moreover, their analysis revealed that
three-dimensional perturbations are accompanied by circulating horizontal mean flows
at large distances from the vicinity of the beam.

They studied the linear stability of uniform internal wave beams with confined
spatial profile 
 by introducing infinitesimal disturbances to the basic state, in the form of normal modes,
not only in the along-beam direction~$\xi$
but also in the horizontal transverse direction~$y$ (see \textbf{Figure~\ref{profilselonetabb}}).
 They used an asymptotic approach, valid for long wavelength
perturbations relative to the beam thickness.
The boundary conditions combined with the matching conditions between the solution near and far from the beam
 ensure that the primary-harmonic and mean-flow perturbations are
 confined in the cross-beam direction.

The analysis brings out the coupling of the primary-harmonic and mean-flow
perturbations to the underlying internal wave beam: 
the interaction of the primary-harmonic perturbation with the beam induces a
mean flow, which in turn feeds back to the primary harmonic via interactions with
the beam. Whether this primary-harmonic-mean flow interaction
mechanism can extract energy from the basic beam, causing instability, depends upon
finding modes which remain confined in
the cross-beam direction.

\subsubsection{Complete model for the 3d 
 propagation of small-amplitude internal wave beams}\label{completemodel}

 In a second stage, \cite{KataokaAkylas2015} have derived a complete matched asymptotic analysis of the experiment
performed by~\cite{Bordes2012}  for a 3D Boussinesq weakly nonlinear viscous fluid uniformly stratified.
From their prior experience \citep{Akylas2003,KataokaAkylas2013}, they have chosen stretched  along-beam
spatial coordinate as ${\Xi}=\varepsilon^2 \xi$, 
 slow time as 
 $T=\varepsilon^2 t$
and  transverse variations as $Y=\varepsilon y$
so that along-beam and transverse dispersions 
 are comparable, together with variations in the cross-beam direction~$\eta$ (see \textbf{Figure~\ref{profilselonetabb}}). 
Combining this choice with a small nonlinearity 
{scaling as $\varepsilon^{1/2}$} and a weak viscous dissipation  
${\bar\nu} \varepsilon^{2}$ that carry equal weight, they were able to fully analyze the mean flow,
 separately near and far from the beam,
before matching both solutions. 

They derived a closed system of two coupled equations linking the amplitude of the primary time harmonic ${\cal U}$ and
the mean-flow component $\overline V_\infty$ of the cross-beam velocity field.
The latter appears to be necessary for matching 
with the mean flow far from the beam. 
The equation governing the dynamics of the mean flow reads
\begin{equation}
\partial_T \overline V_\infty = \cos\theta\, \partial_Y {\mathcal H} \left(\int_{-\infty}^{+\infty} \mbox{d} \eta  \ {\cal U}^* \left(
\frac{\partial \cal U}{\partial{\Xi}}+
\frac{\cot\theta}{2} \int^{\eta} \mbox{d} \eta  
\,\frac{\partial^2 \cal  U}{\partial{Y^2}}\right)\right),\label{equationformeanflowinit}
\end{equation}
 where ${\mathcal H}(.)$ stands for the Hilbert transform in the transverse coordinate $Y$. 
This  
immediately shows that transverse variations $(\partial_Y\neq0)$ of the beam are essential for having a nonzero source term.  

Since the generated mean vertical vorticity 
 is given at leading order by $\overline \Omega
 =\cos\theta \, \partial_Y\overline {\cal U}=(\cos^2\theta/\sin\theta) \, \partial_Y {\overline V_\infty }+{O(\varepsilon^{1/2})}$, 
a direct comparison with Equation~(\ref{equationmeanflowGuilhem}) is possible. 
The first term in Equation (\ref{equationformeanflowinit}), which involves derivatives in both horizontal coordinates corresponds
to the term identified by~\cite{Bordes2012}, while this more complete analysis sheds light on an additional term deriving from purely transverse variations.

Using an intermediate equation,  \cite{KataokaAkylas2015}  end finally with the alternative and more elegant  form
\begin{equation}
\partial_T \overline V_\infty = i \partial_Y {\mathcal H} \left(\int_{-\infty}^{+\infty} \mbox{d} \eta  \left[ \left({\cal U}^*\partial_\eta {\cal U}\right)_T+{\bar\nu}\partial_\eta {\cal U}^*\partial_{\eta\eta} {\cal U}\right]\right).\label{equationformeanflow}
\end{equation}
Moreover, they show that to match inner and outer solutions, this induced mean flow turns out to be purely horizontal to leading order
 and also dominant over the other harmonics. 
The comparison of this theoretical description agrees very well with the experimental results as beautifully emphasized by the 
different panels presented in \textbf{Figure~\ref{ResultmanipGuilhem}}.

As far as a comparison with the experimental results of \cite{Bordes2012}
is concerned, a common caveat of the predictions by \cite{Bordes2012}
and \cite{KataokaAkylas2015} is the assumption of small wavelength
compared to the length scale of the wave envelope, which is only marginally
satisfied in the experiments. One may for instance wonder if the horizontal
structure of the observed waves is primarily due to the feedback of
the mean flow on the wave, or to the sole diffraction pattern of the
wave due to this absence of scale separation. This  needs to be
addressed in future works.

\subsection{Forcing of Oceanic Mean Flows}

Using an analysis based on the Generalized-Lagrangian-Mean (GLM) theory,
\cite{GrisouardBuhler2012} have also studied the role of dissipating oceanic internal
tides in forcing mean flows. {For analytical convenience, they model wave dissipation as a linear damping term  $-\gamma_b b$ in the buoyancy equation (\ref{eq:cons_masse}), and neglect the viscous term in the momentum equation~(\ref{eq:NS_strat}).} 

Within this framework, they discuss in detail the range of situations in which a strong, secularly growing mean-flow response can be expected. Their principal results include the derivation of an expression for the effective mean force exerted by small-amplitude internal tides on oceanic mean flows. At leading order, taking into account the background rotation and using a perturbation series in small wave amplitude, they derive the following explicit expression
 \begin{equation}
\partial_t\overline \Omega+\frac{\gamma_b f}{N^2}\partial_z \overline b=\frac{-i{\gamma_b}N^2}{2(\omega^2+\gamma_b^2)\omega}
\left({\mbox{\boldmath $\nabla$} u_z^*\times\mbox{\boldmath $\nabla$} u_z}\right) \cdot {\mbox{\boldmath $e$}_z},\label{EquationbulhlerGrisouard}
\end{equation}
for the average over the tidal period of the vertical vorticity. It is remarkable that one recovers in the presence of rotation a forcing term on the right-hand side that is analogous to the forcing terms  obtained by \cite{Bordes2012} and \cite{KataokaAkylas2015} in the non-rotating case. In inviscid rotating flows, vortical modes are at geostrophic equilibrium, and there is a frequency gap separating those geostrophic modes from inertia-gravity waves. This frequency gap generally precludes interactions between geostrophic modes and wave modes. The work of  \cite{GrisouardBuhler2012}, however, shows that the combination of nonlinear and dissipative effects allows for a one-way energy transfer from inertia-gravity wave modes to geostrophic modes, through a genuinely three dimensional mechanism.
 {Using} Equation~(\ref{EquationbulhlerGrisouard}), \cite{GrisouardBuhler2012}   compute the effective mean force numerically in a number of idealized examples with simple topographies. 

Although a complete formulation with dissipative terms in the momentum equation is necessary,
the conclusion of this important work by \cite{GrisouardBuhler2012} is that energy {of inertia-gravity waves in rotating fluids }can be transferred to a horizontal mean flow by a similar resonance mechanism as described in the experiment by \cite{Bordes2012}. One understands therefore that mean flows can be generated in regions of wave dissipation,  and not necessarily near the topographic wave source.

\section{CONCLUSIONS AND FUTURE DIRECTIONS} \label{ConclusionsPerspectives}

We have presented several recent experimental and theoretical works that have 
renewed the interest of internal 
wave beams.  After emphasizing the reason for their ubiquity in stratified fluids  -- they are solutions
of the nonlinear governing equations -- this review has presented the two main mechanisms of
instability for those beams:

i) Triadic Resonant Instability. We have shown that this instability produces a 
direct transfer of energy from large scales (primary waves) to smaller scales (subharmonic ones)
 for inviscid plane waves, but that it is no longer true for internal wave beams since the most unstable 
triad may combine subharmonic waves with larger and smaller wavelength.
Moreover, 
 the effects of the finite size
and envelope shape for the onset of Triadic Resonant Instability have been overlooked. These features have to be taken into account
to safely reproduce the complete nonlinear transfer of energy between scales 
in the ocean interior or in experimental analog~\citep{SED2013,BrouzetEPL2016}, and therefore to find its stationary state, 
the so-called Garrett and Munk Spectrum~\citep{GarrettMunk1975}
or its possible theoretical analog, the Zakharov spectrum for the wave turbulence theory~\citep{NazarenkoBook}.

ii) Streaming Instability.
Now that the  mechanism underlying streaming instability and the conditions for its occurrence have been identified,  several other examples will probably be reported in the coming years.
For example, such a mean-flow generation has also been observed in a recent experiment~\citep{Brouzetetal2016}
for which the reflection of internal gravity waves in closed domains lead to an
internal wave attractor. Two lateral Stokes boundary
layers generate indeed a fully three-dimensional interior velocity field that provides the condition for the mean flow to appear.
With a perturbation approach, \cite{Beckebanze2016} confirmed this theoretically and showed that the generated 3D velocity field damps the wave beam at
high wave numbers, thereby providing a new mechanism to establish an energetic balance
for steady state wave attractors. 
\cite{SeminFauve}  have also recently studied experimentally the generation of a
mean flow by a progressive internal gravity wave in a
simple two-dimensional geometry, revisiting an experimental analog of the quasi-biennial 
oscillation~\citep{PlumbMcEwan1978}.
They study the feedback of the mean
flow on the wave, an essential ingredient of
the quasi-biennial oscillation. 

Which is the dominant mechanism? 
\cite{KataokaAkylas2016} have recently suggested that streaming instability are central to three-dimensional 
internal gravity wave beam dynamics in contrast to the TRI of sinusoidal wave train relevant to uniform beams, the special case of a internal plane  wave with confined
spatial profile.
This review reinforces therefore the need for more three-dimensional experiments studying wave-induced mean flow.
In particular, the conditions that favor mean-flow generation with respect to triadic resonant interaction remains
largely unknown. Angles of propagation? Three-dimensionality? Stratification? This is an important question that
needs to be addressed.

\begin{summary}[SUMMARY POINTS]
In an incompressible non-rotating linearly stratified Boussinesq fluid,
\begin{enumerate}
\item Plane waves are solutions of the linear and nonlinear equations for any amplitude. 

\item Internal wave beams, which correspond to the superposition of plane waves with wave vectors of different magnitude but pointing in the same direction, are solutions of the linear and nonlinear equations. 

\item Plane waves solutions are always unstable by TRI.

\item General localized internal wave beams are stable 
 while (quasi) spatial-harmonic internal wave beams are unstable if the 
  beam is wide enough.

\item In presence of rotation, beams of general spatial profile are more vulnerable to TRI
especially close to the critical latitude where nearly-stationary wavepackets remain in the interaction 
region for extended durations, facilitating energy transfer.

\item Internal gravity wave beams with confined
spatial profile are linearly unstable to three-dimensional modulations.

\item When the wave beam is attenuated along its direction of propagation and when the wave-envelop varies in the transverse horizontal direction, nonlinear interactions of the wave beam with itself induce the emergence of a horizontal mean-flow with vertical vorticity.

\end{enumerate}
\end{summary}

\section*{DISCLOSURE STATEMENT}
The authors are not aware of any biases 
 that might be perceived as affecting the objectivity of this review. 

\section*{ACKNOWLEDGMENTS}
This work was supported by the LABEX iMUST (ANR-10-LABX-0064) of Universit\'e de Lyon, within the program "Investissements d'Avenir" (ANR-11-IDEX-0007) operated by the French National Research Agency (ANR). This work has 
achieved thanks to the resources of PSMN from ENS de Lyon.
We acknowledge the contributions of G. Bordes, B. Bourget, C. Brouzet, P.-P. Cortet, E. Ermanyuk, M. Le Bars, 
P. Maurer, F. Moisy, J. Munroe, H. Scolan, and A. Wienkers to our research on this topic.
We thank T. Akylas, V. Botton, N. Grisouard, H. Karimi, T. Kataoka, L. Maas, T. Peacock, C. Staquet, B. Voisin for helpful discussions.

\end{document}